\documentclass{aptpub}
\usepackage{graphicx}
\usepackage{color}
\usepackage{subfigure}

\usepackage{amsfonts}
\usepackage{amsmath}

\newcommand{\ave}[1]{\left \langle #1 \right \rangle}

\newcommand{\I}{\mathcal{I}}
\newcommand{\Sus}{\mathcal{S}}
\newcommand{\PE}{\mathcal{P}}
\newcommand{\A}{\mathcal{A}}
\newcommand{\order}{\mathcal{O}}

\newcommand{\qed}{\hfill$\square$}

\authornames{Joel C. Miller}
\shorttitle{Epidemic Size and Probability}

\newcommand{\Expected}{\mathbb{E}}
\renewcommand{\PE}{\mathcal{Y}}

\begin{document}
\title{Bounding the Size and Probability of Epidemics on Networks}
\author[Division of Mathematical Modeling, British Columbia Centre for
Disease Control\footnote{Much of this work was completed while at the Center for Nonlinear Studies and the 
Mathematical Modeling \& Analysis Group, Los Alamos National
Laboratory.}]{Joel C. Miller}
\address{655 W 12th Ave, Vancouver, BC V5Z 4R4, Canada}

\begin{abstract}
  We consider an infectious disease spreading along the edges of a
  network which may have significant clustering.  The individuals in
  the population have heterogeneous infectiousness and/or
  susceptibility.  We define the \emph{out-transmissibility} of a node
  to be the marginal probability that it would infect a randomly
  chosen neighbor given its infectiousness and the distribution of
  susceptibility.  For a given distribution of out-transmissibility,
  we find the conditions which give the upper [or lower] bounds on
  size and probability of an epidemic, under weak assumptions on the
  transmission properties, but very general assumptions on the
  network.  We find similar bounds for a given distribution of
  in-transmissibility (the marginal probability of being infected by a
  neighbor).  We also find conditions giving global upper bounds on
  size and probability.  The distributions leading to these bounds are
  network-independent.  In the special case of networks with high
  girth (locally tree-like), we are able to prove stronger results.
  In general the probability and size of epidemics are maximal when
  the population is homogeneous and minimal when the variance of in-
  or out-transmissibility is maximal.
\end{abstract}

\keywords{Epidemiology, Networks, Attack Rate, Probability, Transmissibility}

\ams{92D30}{60K35}

\section{Introduction}
\label{sec:intro}

The spread of infectious disease is governed by many different
factors which vary on the individual level.  Heterogeneity in the
population comes from a number of sources including, but not limited
to, genetic diversity, previous infections, vaccination history, or
existence of co-infections.  In this paper we investigate the effects
of heterogeneity on disease spread, focusing on the effect of
simultaneous heterogeneities in infectiousness and susceptibility.

\begin{figure}
\subfigure{\includegraphics[width=0.3\textwidth]{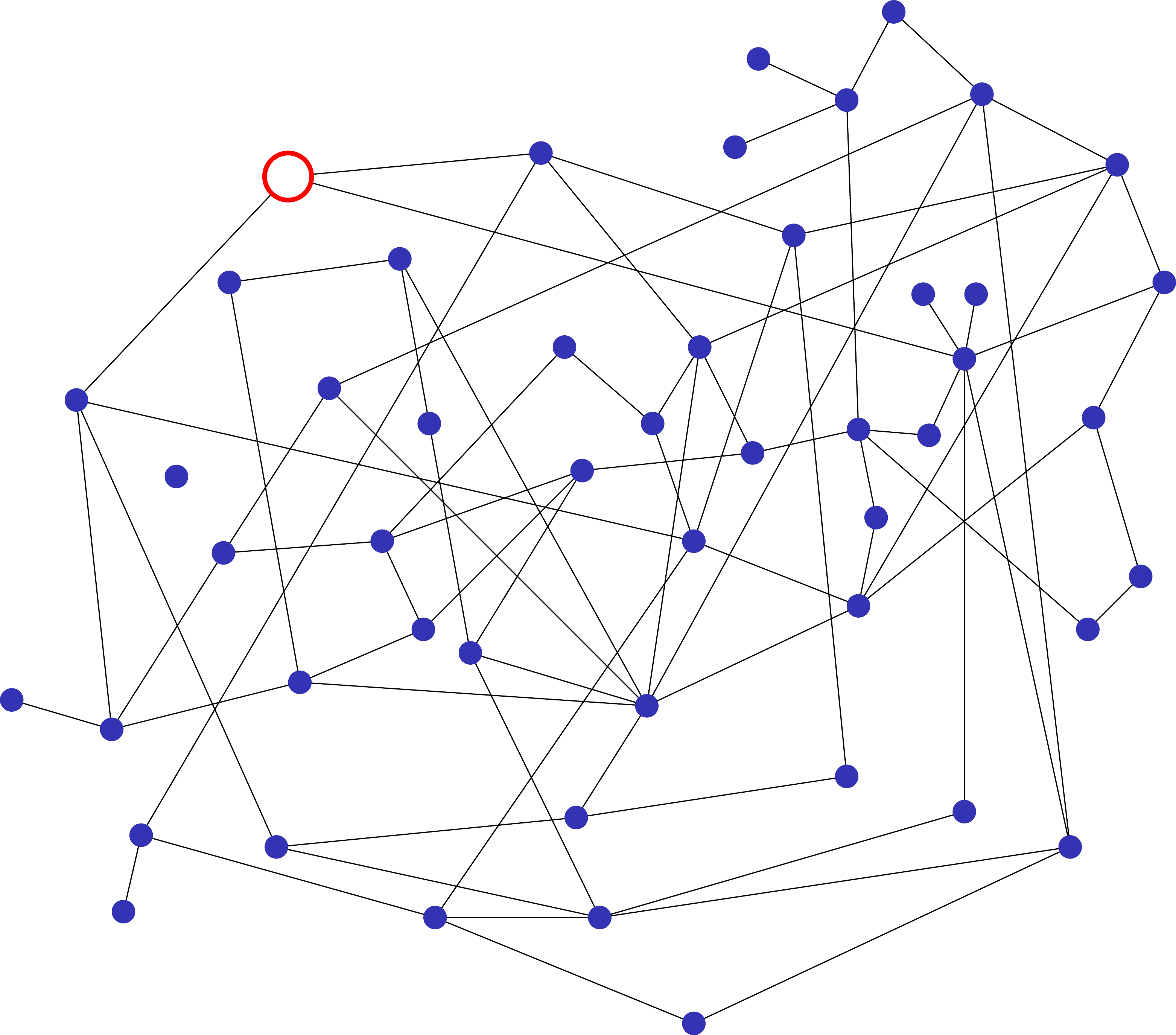}} \hfill
\subfigure{\includegraphics[width=0.3\textwidth]{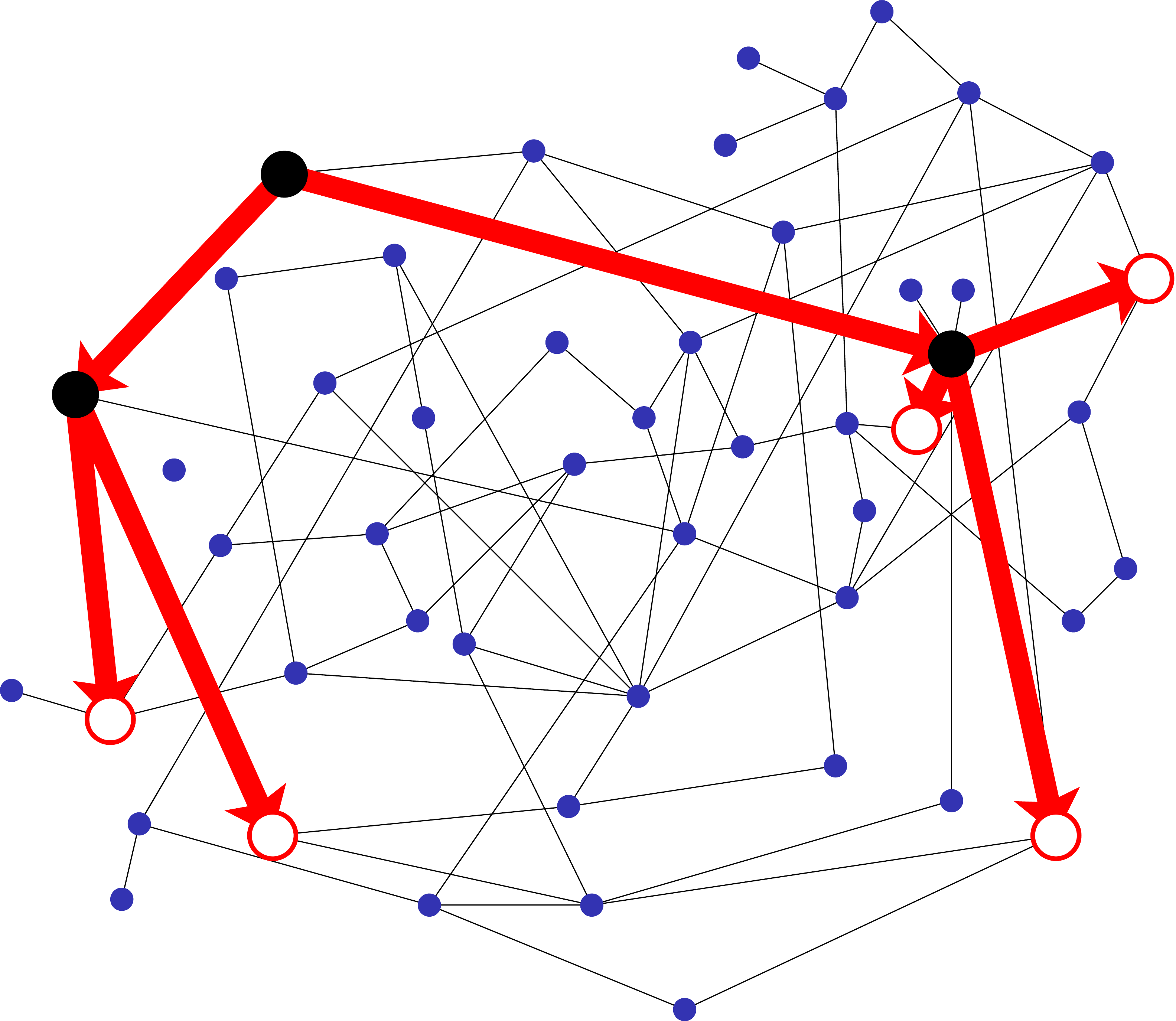}} \hfill
\subfigure{\includegraphics[width=0.3\textwidth]{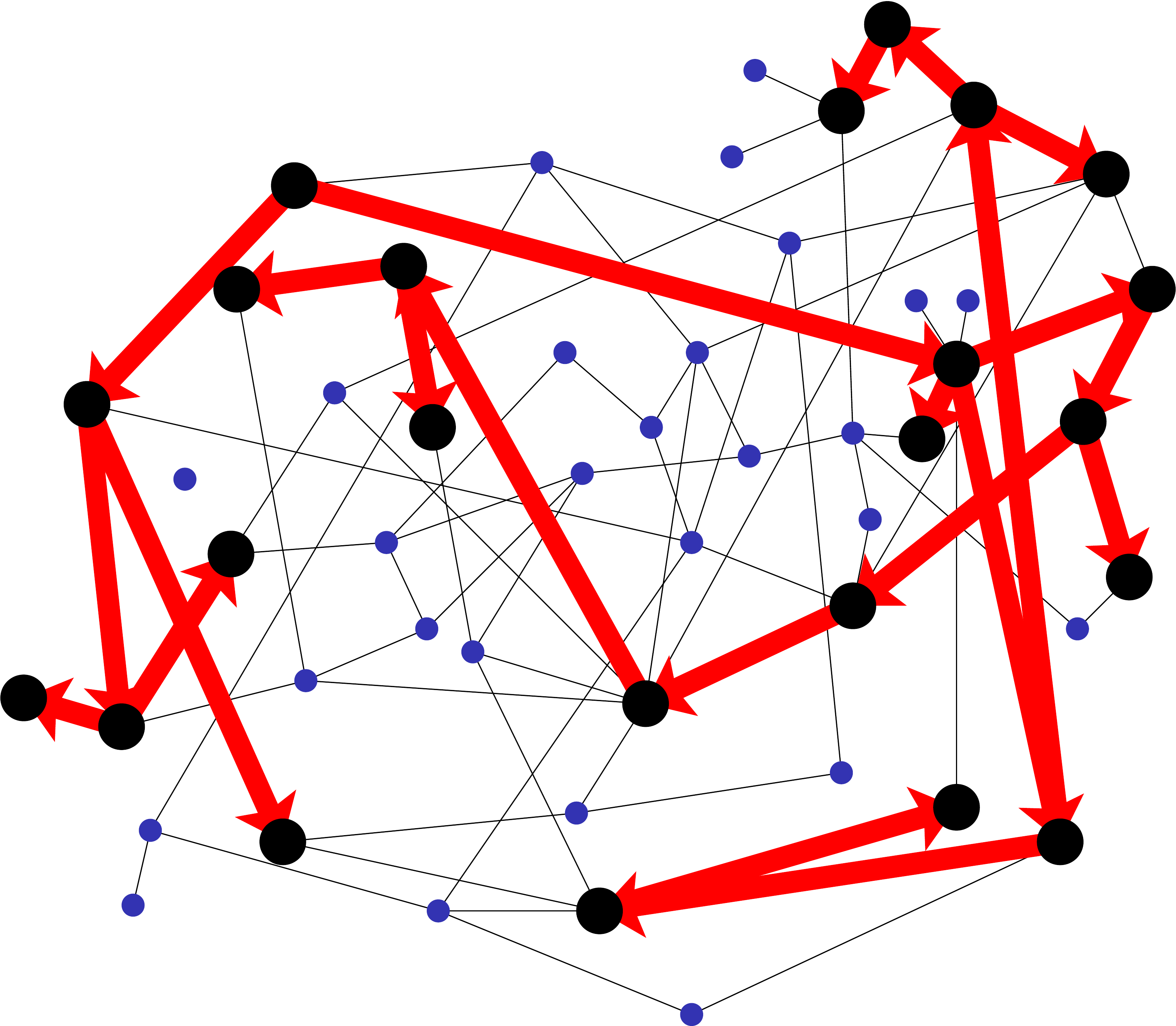}}
\caption{The spread of disease in a network.  The outbreak begins with
  a single infected individual [large empty circles] and then spreads
  along edges to others.  The infected nodes recover with immunity
  [large filled circles].  Eventually the outbreak dies out.}
\label{fig:samplespread}
\end{figure}

We consider the spread of infectious diseases in networks as shown in
figure~\ref{fig:samplespread}.  Individuals in the population are
modeled as nodes and potentially infectious contacts are modeled as
edges between the corresponding nodes.  We consider the spread of an
SIR disease, that is the nodes are divided into three compartments:
\emph{Susceptible}, \emph{Infected}, and \emph{Recovered}.  A
susceptible node may be infected by an infected neighbor.  Following
infection, the newly infected node may infect some, all, or none of its
neighbors and then recover.  After recovery, a node cannot be
reinfected.  Typically in a large network outbreaks are either small
or large (in a sense made more formal in section~\ref{sec:model}).  We
are primarily interested in what controls the probability of large
outbreaks and the fraction of nodes infected in a large outbreak.

Before discussing earlier results, we introduce some terminology.
The \emph{transmissibility} $T_{uv}$ is the probability that an
infection of node $u$ would result in direct infection of the neighbor
$v$.  The \emph{in-transmissibility} $T_{in}(v)$ is the
marginal probability that a neighbor of $v$ would infect $v$ given the
characteristics of $v$, and the \emph{out-transmissibility}
$T_{out}(u)$ is the marginal probability that $u$ would infect a
neighbor given the characteristics of $u$.  Both the in- and
out-transmissibility necessarily have the same average,
$\ave{T}$.  These definitions will be made more
precise in section~\ref{sec:model}.

Most network-based epidemic models assume homogeneous transmissibility
$T_{uv} = \ave{T}$ between all pairs of neighboring nodes.  Models
that do allow heterogeneities generally show that they reduce the
probability or size of
epidemics~\cite{ball:susceptibles,kuulasmaa:locallydep,trapman,miller:heterogeneity,kenah:second}.
For an arbitrary network with homogeneous susceptibility [$T_{in}(v) =
\ave{T}$ for all $v$], but heterogeneous infectiousness,
\cite{kuulasmaa:locallydep} showed that epidemics are most likely and
largest if infectiousness is homogeneous [$T_{out}(u)=\ave{T}$ for all
$u$].  It was noted by~\cite{trapman} that the same argument shows
that with homogeneous susceptibility epidemics are least likely and
smallest if infectiousness is maximally heterogeneous ($T_{out}= 0$
for a fraction $1-\ave{T}$ of the population and $T_{out}=1$ for the
remainder).  The recent work
of~\cite{miller:heterogeneity,kenah:second} considered the effect of
heterogeneity on a specific subclass of unclustered networks
(variously called Molloy--Reed networks~\cite{molloyreed} or
Configuration Model networks~\cite{newman:structurereview}), finding
similar results.  One of these, \cite{miller:heterogeneity}, studied
simultaneous heterogeneities in susceptibility and infectiousness,
showing that for given $\ave{T}$, the same cases give the upper and
lower bounds on probability while epidemics are largest if
susceptibility is homogeneous and smallest if susceptibility is
maximally heterogeneous.  We are unaware of any work which has
considered simultaneous heterogeneities in infectiousness and
susceptibility in networks with clustering or even in unclustered
networks more general than Molloy--Reed networks.

In this paper we investigate the spread of epidemics in which $T_{in}$
and $T_{out}$ can simultaneously be heterogeneous, using techniques
from~\cite{kuulasmaa:locallydep} and~\cite{miller:heterogeneity}.  We
will consider both clustered and general unclustered networks.
Clustered networks are more difficult because of the existence of
short cycles, and so a stronger assumption will be made for them.  

Often only the distribution of $T_{out}$ (or,
more rarely, of $T_{in}$) would be available early in an outbreak.  If
we know the distribution of $T_{out}$, it does not in general uniquely
determine the distribution of $T_{in}$ and so we cannot fully predict
the final details of an outbreak.  Our focus is on identifying the
best and worst case scenarios given the distribution of $T_{out}$ (or
$T_{in}$), thus helping to provide policy makers with knowledge of
what to expect and how best to mitigate it.

Mathematical theories modeling the spread of infectious diseases have
been developed in a number of
fields~\cite{kermack,abbey:reedfrost,andersonMay,eubanks:episims}.
The techniques used include differential equations, stochastic models,
agent-based simulations, and network-based approaches.  The
differential equations approaches may be thought of as a mean-field
approximation to a subclass of network models, while the stochastic
and agent-based approaches can be made formally equivalent to
network-based methods~\cite{kenah:second,neal:coupled}.  Consequently
results for networks will apply to other models as well.  Network
epidemic models have primarily been studied by the statistics
community~\cite{berg:spatial,kuulasmaa:locallydep,kuulasmaa:bondperc,ball:susceptibles,ball:finaldist,mollison,andersson:network}
and the statistical physics/applied mathematics
communities~\cite{newman:spread,meyers:directed,meyers:contact,serrano:prl,hastings:series,pastor-satorras:scale-free,madar,keeling:networkstructure,keeling:networkepidemic}.
In general the statistics community has produced more rigorous
results, but has considered more restricted classes of networks.  The
physics and applied mathematics communities have considered a wider
range of networks, but the results are less rigorous.  The interaction
between these fields has been relatively sparse, leading to repeated
discoveries of some results and a lack of cohesion in the topics
studied.  We attempt to bring some of these different approaches
together in this paper.

This paper is structured as follows: in section~\ref{sec:model} we
introduce the model and clarify definitions.  In
section~\ref{sec:general} we consider epidemics spreading on general
networks.  In section~\ref{sec:unclustered} we find stronger results
for networks with no short cycles.  Finally in
section~\ref{sec:conclusions} we discuss extensions and implications of
our results.

\section{Epidemics in networks}
\label{sec:model}

We consider the spread of disease on a network $G$.  An
\emph{outbreak} begins when a single node (the \emph{index case})
chosen uniformly from the population is infected.  The disease spreads
from an infected node $u$ to a neighboring susceptible node $v$ with a
probability equal to the \emph{transmissibility} $T_{uv}$.  Each
infected node attempts to infect each of its neighbors and then
recovers (and is no longer susceptible or infected).  The outbreak
ends when no infected nodes remain.

This section, like Gaul, is divided into three parts.  First we
describe the neighbor-to-neighbor transmissibility $T_{uv}$.  This
will depend on the characteristics of both $u$ and $v$.  We then
introduce the concept of an Epidemic Percolation Network, which is a
tool to study the routes of transmission in a given network.  We
finally discuss tools which will be used to make the concept of a
``large network'' rigorous.

\subsection{Transmissibility}
\label{sec:trans}

Following~\cite{miller:heterogeneity}, we assume that the factors
influencing infectiousness of node $u$ and susceptibility of node $v$
may be summarized in $\I_u$ and $\Sus_v$.  In general, these may be
vector-valued functions (though with few exceptions they are taken to
be scalars in the literature).  For example, $\I_u$ may represent
$u$'s viral load, duration of infection, and willingness or ability to
leave work if sick, while $\Sus_v$ may represent $v$'s previous
vaccination history, genetic predisposition to infection, and previous
exposure to related infections.  If $u$ and $v$ are neighbors, the
transmissibility $T_{uv}$ is then
\[
T_{uv} = T(\I_u, \Sus_v)
\]
for some function $T$.  The function $T(\I,\Sus)$ is the probability of transmission from a node with infectiousness $\I$ to a node with susceptibility $\Sus$ assuming that the nodes are joined by an edge.  We may think of $T_{uv}$ as defined only for neighboring nodes, or we may take $T_{uv} = \chi_{\{u,v\}} T(\I_u,\Sus_v)$ where $\chi_{\{u,v\}}=0$ if $\{u,v\}$ is not an edge and $1$ if it is.

We assume that $\I$ and $\Sus$ are assigned independently, using the
probability density functions $P(\I)$ and $P(\Sus)$ (although we use
the same symbol $P$ for both, we assume that the two functions are
different).  Particularly if $\I$ and $\Sus$ are vectors, we may not
be able to clearly define which of two nodes is ``more infectious''
(\emph{i.e.}, there may not be a well-defined ordering).  For example,
with a sexually transmitted disease, we might have $u_1$ and $u_2$
infected, with $u_1$ having a high viral load and regular condom use
(with occasional lapses), while $u_2$ has a low viral load but no
condom use.  Let us assume they have contacts with susceptibles $v_1$
and $v_2$ where $v_1$ has a high level of resistance, and thus will
only be infected by a large dose, while $v_2$ has no immune protection
and thus will be infected by even a small dose.  Under these
assumptions, $u_1$ is more likely to infect $v_1$, while $u_2$ is more
likely to infect $v_2$.  Which is ``more infectious'' depends on the
test susceptible considered.

The probability that $u$ infects a neighbor (prior to knowing $\Sus$ for that neighbor) is given by the \emph{out-transmissibility} of $u$
\[
T_{out}(u) = \int T(\I_u, \Sus) P(\Sus) d\Sus
\]
and the probability that $v$ would be infected by a
neighbor is given by the \emph{in-transmissibility} of $v$
\[
T_{in}(v) = \int T(\I, \Sus_v) P(\I) d\I
\]
At times it will be convenient to use $T_{out}(\I)$ and $T_{in}(\Sus)$
[rather than $T_{out}(u)$ and $T_{in}(v)$] to denote the out- and
in-transmissibility of arbitrary nodes with $\I$ and $\Sus$
respectively.  When the concepts of being ``more infectious'' and
``more susceptible'' are clearly defined, $T_{in}(\Sus)$ and
$T_{out}(\I)$ are invertible functions.  However, because the ordering
is not well-defined in general, they may not be invertible.  If they
are invertible, it is often convenient to change variables and set $\I_u = T_{out}(u)$
or $\Sus_v = T_{in}(v)$.  We will do this frequently in
section~\ref{sec:general}, where we restrict our attention to cases where the
ordering described above is well-defined.

From $P(\Sus)$ and $P(\I)$, we may find the distributions of $T_{in}$
and $T_{out}$.  We use $Q_{in}(T_{in})$ to denote the probability
density function for the in-transmissibility $T_{in}$ and
$Q_{out}(T_{out})$ to denote the probability density function for the
out-transmissibility $T_{out}$.  The averages $\int_0^1 T_{in}
Q_{in}(T_{in})dT_{in}$ and $\int_0^1 T_{out}Q_{out}(T_{out})dT_{out}$ are
both equal to $\ave{T}$.

Given distributions of $\I$ and $\Sus$ and the function $T$, there is
always a $Q_{in}$ and $Q_{out}$ pair that result.  Also, given a
$Q_{in}$ or a $Q_{out}$ it is always possible to find $P(\I)$,
$P(\Sus)$, and $T$ that are consistent.  For example, given any
$Q_{in}$, for each node $v$ we assign a $T_{in}(v)$ from $Q_{in}$ and
set $\Sus_v = T_{in}(v)$.  Then $T(\I,\Sus) = \Sus$ is consistent with
$Q_{in}$ and yields $Q_{out}(T_{out}) = \delta(T_{out}-\ave{T})$.
This means that for any distribution of in-transmissibility, it is
possible that the infectiousness of nodes is homogeneous.  Although it
is possible to find a $Q_{in}$ for any $Q_{out}$ (and \emph{vice
  versa}) not all pairs $Q_{out}$ and $Q_{in}$ are compatible.  For
example, if $Q_{in}=(1-\ave{T})\delta(T_{in}) +
\ave{T}\delta(T_{in}-1)$ (\emph{i.e.}, susceptibility is maximally
heterogeneous) then the out-transmissibility must be homogeneously
distributed; no other distribution is possible.  This particular
example will be important in Section~\ref{sec:unclustered}.

Although in principle $\I$ and $\Sus$ may be vector-valued, they
frequently are assumed to be scalars with the transmissibility between
two neighbors given by (for
example~\cite{valle:mixing,miller:heterogeneity})
\begin{equation}
T_{uv} = T(\I_u,\Sus_v) = 1 - \exp (-\alpha \I_u \Sus_v)
\label{eqn:exptrans}
\end{equation}
A number of disease models yield this form.  For example: let
$\alpha$ be the rate at which virus from an infected person reaches a
susceptible person.  Let $\I_u$ be the infectious period of $u$.  Let
$\Sus_v$ be the probability that a virus reaching $v$ causes
infection.  Then the probability $p$ that $v$ has not become infected
satisfies
$\dot{p} = - \alpha \Sus_v p$.
Integrating this over the infectious period $\I_u$ of $u$
yields equation~\eqref{eqn:exptrans}.

We need one final concept related to the transmissibility.  Let a node $u$ be given, and let $V = \{v_1, \ldots, v_m\}$ be a subset of the neighbors of $u$.  Assume we know $\vec{\Sus} = (\Sus_{v_1}, \ldots, \Sus_{v_m})$, but not $\I_u$.  Define 
\begin{equation}
\phi_{in}(V,\vec{\Sus}) = \int \prod_{v \in V} [1-T(\I,\Sus_v)] P(\I)
d\I \, .
\label{eqn:phiV}
\end{equation}
This is the probability that $u$ will not infect any node in $V$ given
knowledge of $\Sus$ for each $v \in V$, but marginalized over the
possible values of $\I$ for $u$.  We may similarly define
\begin{align}
\psi_{in}(V) &= \int \phi_{in}(V, \vec{\Sus}) P(\vec{\Sus}) d\vec{\Sus} \, ,\label{eqn:psiphi}\\
   &= \int (1-T_{out})^{|V|} Q_{out}(T_{out}) dT_{out} \, .\label{eqn:psiV}
\end{align}
This is the probability that $u$ will not infect any $v\in V$
marginalized over $\Sus$ of $v \in V$ and the values of $\I$ for $u$.
If $|V|=1$, then $\psi_{in}(V) = 1-\ave{T}$, which will be important
later when we consider unclustered networks.


\subsection{Epidemic Percolation Networks}
\label{sec:EPN}

Given a network $G$, the distributions $P(\I)$ and $P(\Sus)$, and the
function $T(\I,\Sus)$, we assign $\I$ and $\Sus$ to each node of $G$.
We then create a new directed network $\mathcal{E}$ which is an
\emph{Epidemic Percolation Network} (EPN)~\cite{kenah:networkbased} as follows: the nodes of
$\mathcal{E}$ are the nodes of $G$.  For each edge $\{u,v\}$ of $G$,
we place directed edges $(u,v)$ and $(v,u)$ into $\mathcal{E}$ with
probability $T_{uv}$ and $T_{vu}$ respectively.  The original network
$G$ gives the paths a disease \emph{could} follow, while a
realization of $\mathcal{E}$ gives the paths the disease \emph{will}
follow (if given the chance) for a simulation.

The out-component of a given node $u$ found by assigning $\I$ and
$\Sus$ and generating an EPN comes from the same distribution as the
nodes infected by the dynamic epidemic process described earlier with
$u$ as the index case.  The processes are formally equivalent.

To motivate some definitions, we assume sufficiently high
transmissibility that there are nodes in $\mathcal{E}$ with giant in-
or out-components~\cite{broder}.  We define $H_{out}$ to be those
nodes with a giant in-component, and $H_{in}$ to be those nodes with a
giant out-component in $\mathcal{E}$.  We define $H_{scc} = H_{in}
\cap H_{out}$.  $H_{scc}$ will almost surely be a strongly connected
component.  $H_{in}$ is the in-component of $H_{scc}$ and $H_{out}$ is
its out-component.  In general, infection of any $u \in H_{in}$
results in infection of all nodes in $H_{out}$ and occasionally a few
other nodes (if $u \not \in H_{scc}$).  We define such an outbreak to
be an \emph{epidemic}.  If $u \not \in H_{in}$, then a small
\emph{self-limiting} outbreak occurs.

For large values of $N = |G|$, the probability of an epidemic is given
by $\PE = \Expected[|H_{in}|]/N$ and the expected fraction infected in
an epidemic is given by $\A=\Expected[|H_{out}|]/N + \order\left(\log
  N/N\right)$.  As $N$ grows, $|H_{out}|/N$ approaches
$\Expected[|H_{out}|]/N$, and so the size of a single epidemic in a
large population closely approximates expected size of epidemics (note
that if we include non-epidemic outbreaks in the average, this does
not hold).

If the directions of arrows in the EPNs are reversed, then $H_{in}$
and $H_{out}$ interchange roles.  Consequently, replacing $T_{uv} =
T(\I_u, \Sus_v)$ with $\hat{T}_{uv} = T(\I_v, \Sus_u)$ interchanges
the size and probability.  As such, results derived for the
probability of an epidemic also apply to the size.

\subsection{Large Networks}

The results we derive will be appropriate in the limit of ``large
networks''.  However, in practice we are usually interested in a
single given network.  Unfortunately $|G| \to \infty$ is a vague
concept when we are given a single, finite network.  There are many
ways to increase its size, with different impacts on epidemics.  In
this section, we define what is meant by $|G| \to \infty$ in a way
that allows us to produce rigorous results.

Consider a sequence of networks $\{G_n\}$ which satisfy $|G_n| \to
\infty$ as $n \to \infty$.  We define an open ball $B_d(u)$ to be a
network centered at a node $u$ such that all nodes $v \in B_d(u)$ are
at most a distance $d$ from $u$.  Given a network $G$, we define
$P_G(B_d(u))$ to be the probability that if we choose a node $\hat{u}$
randomly from $G$, then the set of nodes of distance at most $d$ from
$\hat{u}$ is isomorphic to $B_d(u)$ (with the isomorphism mapping
$\hat{u}$ to $u$).

We define \emph{sequential convergence of local statistics} to mean
that given any $d$ and $B_d(u)$, $P_{G_n}(B_d(u)) = P_{G_d}(B_d(u))$
for all $n \geq d$.  For the results developed later, all that is
strictly needed is that $P_{G_n}(B_d(u))$ converges as $n\to\infty$,
but the stronger statement that for $n \geq d$ they do not change
makes the proofs simpler.  This means that for large enough $n$,
networks have the same ``small-scale'' structure, and the size of what
is considered ``small-scale'' increases with $|G|$.  We restrict our
attention to sequences which have sequential convergence of local
statistics.

For a given EPN, we define $H_{in}(d)$ and $H_{out}(d)$ to be the set
of nodes from which a path of length (at least) $d$ begins or ends
respectively.  At large $d$, these will correspond to the $H_{in}$ and
$H_{out}$ described earlier.  We define $\PE_d(G)$ and $\A_d(G)$ to be
the probability that a randomly chosen node from $G$ is in $H_{in}(d)$
and $H_{out}(d)$ respectively.  Sequential convergence means that
$\PE_d(G_n) = \PE_d(G_d)$ and $\A_d(G_n)=\A_d(G_d)$ for $n\geq d$.  We
finally define
\begin{align*}
\PE &= \lim_{d \to \infty} \PE_d(G_d) \, ,\\
\A &= \lim_{d\to\infty} \A_d(G_d) \, .
\end{align*}
$\PE$ measures the probability of an epidemic and $\A$ measures the
fraction infected.  

We will prove our results in the limit $n \to \infty$ by showing that $H_{in}(d)$ and $H_{out}(d)$ for a given $G_n$ are maximal or minimal under different conditions.  This means that our results are generally true for arbitrary finite networks.  The reason we use the large $n$ limit is because for networks which are small it is unclear what constitutes a giant component in an EPN, or similarly, for a network with some unusual structure on a size comparable to the network size (for example a network made up of a few disconnected components), a giant component may not be uniquely defined.  Using the large $n$ limit avoids these problems.  We could avoid the need for a limit by instead assuming the existence of a giant strongly connected component in the EPN and showing that the same conditions maximize or minimize the probability a node is in the in- or out-component of this giant strongly connected component.

\section{Bounds in general networks}
\label{sec:general}

We begin by considering the spread of infectious diseases on arbitrary
networks.  We begin with a simple lemma which we will need in this
section and the next.

\begin{lemma}[Edge Reversal]
\label{lem:interchange}
Given $T_{uv}= T(\I_u,\Sus_v)$, if we interchange the roles of
infectiousness and susceptibility so that $T_{uv} = T(\I_v,\Sus_u)$
for all edges, then $\PE$ and $\A$ interchange roles.
\end{lemma}

\begin{proof}
  If we replace $T_{uv} = T(\I_u,\Sus_v)$ with $\hat{T}_{uv} =
  T(\I_v,\Sus_u)$, then the new EPNs correspond to reversing the
  direction of edges in the original EPNs.  Since reversing the
  direction of edges in an EPN interchanges $H_{in}(d)$ and
  $H_{out}(d)$, this interchanges $\PE$ and $\A$, and finishes the proof.
\qed
\end{proof}

We now make a simplifying assumption which we will need for networks
with short cycles.  

\begin{assumption}[Ordering Assumption]
  If $T(\I_1,\Sus_1) > T(\I_2, \Sus_1)$ for any $\Sus_1$, then
  $T(\I_1,\Sus) \geq T(\I_2, \Sus)$ for all $\Sus$.  Further, strict
  inequality occurs for a set of positive measure.  Similarly if
  $T(\I_1,\Sus_1)> T(\I_1,\Sus_2)$ for any $\I_1$, then $T(\I,\Sus_1)
  \geq T(\I,\Sus_2)$ for all $\I$ with strict inequality for a set of
  positive measure.
\end{assumption}

The ordering assumption is a statement about the functional form of
$T(\I,\Sus)$.  It places no restrictions on the network.  The
assumption holds for equation~\eqref{eqn:exptrans}, but as noted
earlier there are many scenarios where it fails.

The ordering assumption implies that $T_{out}(\I)$ and $T_{in}(\Sus)$
are invertible mappings.
It also allows us to assume that $\I$ is a
scalar quantity ordered such that
\[\I_{u} \geq \I_{u'} 
               \quad \Leftrightarrow \quad T(\I_{u},\Sus) \geq T(\I_{u'}, \Sus) \:\:\:
               \forall \Sus
\quad
\Leftrightarrow \quad T_{out}(u) \geq T_{out}(u')
\]
and further $\I_{u} > \I_{u'} \Leftrightarrow T_{out}(u) >
T_{out}(u')$.  We may make similar conclusions about $\Sus$.  There
will be more than one way to represent $\I$ or $\Sus$ as scalars.  It
will frequently (but not always) be convenient to identify $\I$ with
$T_{out}(\I)$ and $\Sus$ with $T_{in}(\Sus)$.  

Previous work by~\cite{kuulasmaa:locallydep} considered the spread of
infectious diseases on networks for which the only heterogeneity came
from variation in duration of infection.  Hence all nodes have the
same $T_{in}$, and variation occurs only in $T_{out}$.  This model
satisfies the ordering assumption.  In this section we generalize the
results of~\cite{kuulasmaa:locallydep} by allowing $T_{in}$ and
$T_{out}$ to be heterogeneous simultaneously.

We will drop the ordering assumption in section~\ref{sec:unclustered}
where we consider networks with no short cycles.  Even in this
section, many of the results hold without the ordering assumption, but
the proofs are less clean.  The assumption is only strictly needed for
Theorems~\ref{thm:OAupper}, \ref{thm:OAupper_lower_interchange}, and
\ref{thm:OAglobal_upper}.

We are now ready to show that increased heterogeneity generally
decreases the size and probability of epidemics.  We show that for a
given $Q_{in}$ [resp.\ $Q_{out}$], both $\PE$ and $\A$ are maximal when
$T_{out}$ [resp.\ $T_{in}$] is homogeneous.  They are minimal when the
variance of $T_{out}$ [resp.\ $T_{in}$] is maximal subject to the constraint of $Q_{in}$ [resp.\ $Q_{out}$].

We can also derive conditions for a global upper bound on $\PE$ and
$\A$.  The upper bounds occur when $T_{uv} = \ave{T}$ for all
neighbors $u$ and $v$.  We hypothesize a lower bound, but cannot prove
it in networks with short cycles.

To make the notation cleaner in the following lemma, we identify $\Sus$
with $T_{in}$ and so we may use $T(\I,T_{in})$ in place of
$T(\I,\Sus)$.

\begin{lemma}
\label{lem:general}
Assume a sequence of networks $\{G_n\}$ with sequential convergence of
local statistics and a susceptibility distribution $Q_{in}(T_{in})$.
Assume the ordering assumption holds and consider a distribution of
infectiousness $P_1(\I)$ with transmissibility given by
$T_1(\I,T_{in})$, that is consistent with $Q_{in}$.  Let
$\phi_{in,1}(V,\vec{\Sus})$ be as in equation~\eqref{eqn:phiV}.  Let
$\A_1$ and $\PE_1$ be the corresponding attack rate and epidemic
probability.  Similarly choose another $P_2(\I)$, $T_2(\I,T_{in})$
with corresponding $\A_2$, $\PE_2$, and $\phi_2(V,\vec{\Sus})$.
Assume that $\phi_{in,1}(V,\vec{\Sus}) \leq \phi_{in,2}(V,\vec{\Sus})$
for all $V$ and $\vec{\Sus}$.  Then $\A_1 \geq \A_2$ and $\PE_1 \geq
\PE_2$.
\end{lemma}

\begin{proof}
  Let $d\geq 0$ be given.

  Take $G_n$, $n \geq d$.  We will show that a node in an EPN created
  from $G_n$ using the first distribution is more likely to be in
  $H_{out}(d)$ than a node in an EPN created using the second
  distribution.

  Choose any node $u$ from $G_n$.  Partition the nodes of $G_n$ into
  disjoint sets $\{u\}$, $U_1$, and $U_2$.  To the nodes in $U_1$ we
  assign $\I$ from $P_1(\I)$ and to the nodes in $U_2$ we assign $\I$
  from $P_2(\I)$.  We assign $T_{in}$ to all nodes from
  $Q_{in}(T_{in})$.  We will consider the effects of adding $u$ to
  $U_1$ versus adding it to $U_2$.

  Consider a partial EPN $\mathcal{E}$ created by assigning edges
  $(w,v)$ from all $w \neq u$, using $T_1(\I_w,T_i(v))$ if $w \in U_1$
  and $T_2(\I_w,T_i(v))$ if $w \in U_2$.  Now consider an arbitrary
  node $u'$ (which may be $u$) which is not already in $H_{in}(d)$,
  but which would join $H_{in}(d)$ if the appropriate edges were added
  from $u$.  Let $V$ be the set of neighbors $v$ of $u$ for which
  adding the edge $(u,v)$ would allow a path from $u'$ to $u$ to be
  extended to a path of length $d$.

  We consider extensions of $\mathcal{E}$ formed by placing $u$ into
  $U_1$ or $U_2$.  The probability that $u'$ would be in $H_{in}(d)$
  in the extended EPN is equal to the probability that $u$ has at
  least one edge to some node in $V$.  This probability is at least as
  high if $u \in U_1$ as if $u \in U_2$ by our assumption
  $\phi_{in,1}(V,\vec{S}) \leq \phi_{in,2}(V,\vec{S})$.  Consequently
  the probability of $u'$ to be in $H_{in}(d)$ is maximal if $u \in
  U_1$.  Induction on $|U_1|$ shows that $\PE_d(G_n)$
  is largest if all nodes are in $U_1$.

  We now show that $u \in U_1$ increases $\A_d$ compared with $u \in
  U_2$.  We can prove that placing $u$ in $U_1$ versus $U_2$ can only
  increase the probability of a node to be at the end of a length $d$
  path.  The proof proceeds largely as above.  Consider the same
  partial EPN $\mathcal{E}$ defined above. Let $u'$ be a node which is
  not in $H_{out}(d)$ but would be if an edge from $u$ to any $v \in
  V$ (note that $u \neq u'$).  The probability that $u'$ will be in
  $H_{out}(d)$ is $\phi_{in,1}(V,\vec{\Sus})$ or
  $\phi_{in,2}(V,\vec{\Sus})$ depending on whether $u$ is assigned
  $\I$ from $P_1$ or $P_2$.  Because $\phi_{in,1}(V,\vec{\Sus}) \leq
  \phi_{in,2}(V,\vec{\Sus})$ it follows that $\A_d$ is largest if $u
  \in U_1$.  Induction on $|U_1|$ shows $\A_d(G_n)$ is maximal if all
  nodes are in $U_1$.

  Taking $d \to \infty$, it follows then that $\PE$ and $\A$ are
  maximal if all nodes are in $U_1$, and so the proof is finished.  \qed
\end{proof}

%

We begin by showing that for fixed distribution of
in-transmissibility, the size and probability are largest when the
out-transmissibility is homogeneous.

\begin{theorem}
  \label{thm:OAupper}
  Let $Q_{in}(T_{in})$ be given. Assume that the ordering assumption
  holds and that $\{G_n\}$ satisfies sequential convergence of
  statistics.  Set $\Sus_v = T_{in}(v)$.  Then $\PE$ and $\A$ are
  maximized when $T(\I,T_{in})=T_{in}$.
\end{theorem}

\begin{proof}
By the ordering assumption, we may take $\I$ to be scalar with $\I_1 > \I_2$ iff $T_{out}(\I_1) > T_{out}(\I_2)$.  This allows us to use Chebyshev's ``other'' inequality~\cite{kingman}: if $h_1$ and $h_2$ are decreasing
  functions of $x$ and $p$ is a probability density function,
  \[
  \int h_1(x) h_2(x) p(x) \, dx \geq \left[ \int h_1(x) p(x) \, dx \right] \left[ \int h_2(x) p(x) \, dx \right]
  \]
  By induction $\int[\prod h_j(x)] p(x) \, dx \geq \prod \int h_j(x) p(x) \, dx$ for any number of decreasing functions $h_j$.

Applying this to the decreasing function $h_j(\I) = 1- T(\I,T_{in}(v_j))$ we have
  \begin{align*}
    \phi_{in}(V,\vec{\Sus}) &= \int \left [\prod_{v\in V} h_j(\I) \right] P(\I) \, d\I\, ,\\
    & \geq \prod_{v\in V}1- T_{in}(v) \, ,
  \end{align*}
  with equality if $T(\I,T_{in})= T_{in}$.  Thus by
  Lemma~\ref{lem:general}, $\A$ and $\PE$ are maximal, completing the proof.
\qed
\end{proof}

We have proven the upper bounds given $Q_{in}(T_{in})$ occur when
$T_{out}$ is homogeneous.  We now show the lower bounds occur when
$T_{out}$ is maximally heterogeneous.  Because of the ordering
assumption, we may take $\Sus_v = T_{in}(v)$.
\begin{theorem}
  \label{thm:OAlower}
  Let $Q_{in}(T_{in})$ be given, assume the ordering assumption holds,
  and assume that $\{G_n\}$ satisfies sequential convergence of
  statistics.  Take $\I$ to be chosen uniformly from $[0,1]$.  Setting
  \begin{equation}
    T(\I,T_{in}) = \begin{cases} 0 & T_{in} < \I\\
                                1 & T_{in} >\I
                              \end{cases}
  \label{eqn:lowerbound}
  \end{equation}
  minimizes $\PE$ and $\A$.
\end{theorem}
\begin{proof}
Given equation~\eqref{eqn:lowerbound}, we have $\phi_{in}(V,\vec{\Sus}) = \min_{v \in V} \{ 1- T_{in}(v)\}$.  

We need to prove that for any arbitrary transmission function
$\hat{T}(\I,T_{in})$ satisfying the ordering assumption and consistent
with $T_{in}$, $\phi_{in}(V,\vec{\Sus}) \leq \min_{v \in V} \{ 1-
T_{in}(v)\}$.  To do this, let $\hat{T}$ be given, $T_{in}$ assigned
to $v_1$, \ldots, $v_n$ and assume $v_1$, \ldots, $v_n$ are ordered
such that $T_{in}(v_1) \geq T_{in}(v_2) \geq \cdots \geq T_{in}(v_n)$.

Then
\begin{align*}
  \phi_{in}(V,\vec{\Sus}) &=\int \prod_{j=1}^n [1-\hat{T}(\I,T_{in}(v_j))] P(\I) d\I \, ,\\
  &\leq \int [1-\hat{T}(\I,T_{in}(v_1)) ] P(\I) d\I \, ,\\
  &\leq 1-T_{in}(v_1) \, .
\end{align*}
This shows that for any $\hat{T}$, $\phi_{in}(V,\vec{\Sus})$ is at
most the value it takes for~\eqref{eqn:lowerbound}.  Thus
Lemma~\ref{lem:general} shows that $\PE$ and $\A$ are minimal, completing the proof.
\qed
\end{proof}

We derived the results above with fixed $Q_{in}$.  Lemma~\ref{lem:interchange} shows that the equivalent results must hold for $Q_{out}$.  

\begin{theorem}
  \label{thm:OAupper_lower_interchange}

  Let $Q_{out}(T_{out})$ be given.  Assume that the ordering
  assumption holds and that $\{G_n\}$ satisfies sequential convergence
  of statistics.  Set $\I_u= T_{out}(u)$.
\begin{itemize}
\item If 
\[
T(T_{out},\Sus) = T_{out}
\]
Then $\PE$ and $\A$ are maximized.
\item If $\Sus$ is chosen uniformly in $[0,1]$ and
\[
T(T_{out}, \Sus) = \begin{cases} 0 & T_{out}<\Sus \\ 1 & T_{out} > \Sus \, ,
\end{cases}
\]
then $\PE$ and $\A$ are minimized.
\end{itemize}

\end{theorem}

\begin{proof}
  This follows immediately from Lemma~\ref{lem:interchange} with
  Theorems~\ref{thm:OAupper} and~\ref{thm:OAlower}.
\qed
\end{proof}

We now give a global upper bound for both $\PE$ and $\A$.
\begin{theorem}
  \label{thm:OAglobal_upper}

  Let $\ave{T}$ be given.  Under the ordering assumption with
  sequential convergence of statistics for $\{G_n\}$, the maximum of
  $\PE$ and $\A$ occur when $T_{uv} = \ave{T}$ for all neighboring
  nodes.
\end{theorem}

\begin{proof}
  Consider a $P(\I)$, $P(\Sus)$, and $T(\I,\Sus)$ which yields a
  global maximum for either $\PE$ or $\A$.  If $T_{in}$ is not
  homogeneous, then we can find a new infection process which
  preserves the same $Q_{out}(T_{out})$ with homogeneous $T_{in}$
  which can only increase $\PE$ or $\A$.  A repeated application
  preserving the new homogeneous in-transmissibility, but now making
  $T_{out}$ also homogeneous again can only increase $\PE$ or $\A$.
  $T_{out}$ and $T_{in}$ are then homogeneous.  This completes the
  proof.\qed
\end{proof}

We finish with a conjecture about global lower bounds.
\begin{conjecture}
  Under the ordering assumption with sequential convergence of
  statistics for $\{G_n\}$, the minimum of $\PE$ occurs when $Q_{out}(T_{out}) =
  \ave{T} \delta(T_{out}-1) + (1-\ave{T})\delta(T_{out})$.

The minimum of $\A$ occurs when $Q_{in}(T_{in}) = \ave{T} \delta(T_{in}-1) +
(1-\ave{T})\delta(T_{in})$.
\end{conjecture}

Note that if $Q_{out}(T_{out}) = \ave{T} \delta(T_{out}-1) +
(1-\ave{T})\delta(T_{out})$, then $Q_{in}(T_{in}) =
\delta(T_{in}-\ave{T})$ is homogeneous.

\subsection{Discussion}

The results of this section have focused on extending earlier results
of Kuulasmaa~\cite{kuulasmaa:locallydep} who considered a population
with homogeneous susceptibility and heterogeneities in
infectiousness due entirely to variation in duration of infection.  We
have extended these results to cover a wide range of heterogeneities
in infectiousness and susceptibility (simultaneously), under the
assumption that infectiousness and susceptibility are assigned
independently.  In order to extend the proof used by Kuulasmaa, we
have been forced to make the ordering assumption, which effectively
means that if we order people by how infectious they would be to one
test susceptible individual, the order is the same as we would find
for another test susceptible individual.  We do not have any
counter-examples to these theorems in the case where the ordering
assumption fails, and so it is not clear that it is needed.  In
section~\ref{sec:unclustered} we will see that similar results hold in
unclustered networks without needing the ordering assumption.

Our results show that in general increasing the heterogeneity of the
population is useful for either decreasing the size or decreasing the
probability that an epidemic occurs.  Given $Q_{in}$ [resp.\
$Q_{out}$], both $\PE$ and $\A$ are maximized if $T_{out}$ [resp.\
$T_{in}$] is homogeneous and minimized if it is maximally
heterogeneous.  Similarly, given just $\ave{T}$, we find that the
global maxima of $\PE$ and $\A$ occur when $T = \ave{T}$.  Perhaps
surprisingly, the conditions leading to upper and lower bounds are
independent of the network, though the size of the variation between
these bounds is network-dependent.

Although we can prove lower bounds given $Q_{in}$ [or $Q_{out}$], we
cannot prove global lower bounds given $\ave{T}$.  We hypothesize that
the global lower bound for $\PE$ occurs when $Q_{out}$ is maximally
heterogeneous and the global lower bound for $\A$ occurs when $Q_{in}$
is maximally heterogeneous.  In the next section we will see that
these are the lower bounds for an unclustered population.  However, we
have not found a rigorous proof for general networks.  In the proof of
the upper bound, we took a given $Q_{out}$ and found $Q_{in}$ that
maximizes $\PE$ and $\A$.  We then held that $Q_{in}$ fixed and found
$Q_{out}$ to maximize, arriving at the upper bound.  However, applying
a similar technique to the lower bound fails because given any
$Q_{out}$, if we find a minimizing $Q_{in}$, attempting to then
minimize with $Q_{in}$ fixed simply returns the original $Q_{out}$.
The difficulty results from the fact that increasing heterogeneity in
$T_{out}$ restricts the amount of heterogeneity in $T_{in}$ and
\emph{vice versa}.

\section{Bounds in unclustered networks}
\label{sec:unclustered}
Most studies of infectious diseases spreading on networks have been
made for networks for which the effect of short cycles may be
neglected~\cite{newman:spread}.  These investigations have generally
used Molloy--Reed networks~\cite{molloyreed} (also known as the
configuration model~\cite{newman:structurereview}).  The theory we
develop here applies to these networks, but also to more general
networks which may have degree-degree correlations, or even longer
range correlations.

When we study networks with no short cycles, we are able to prove
stronger results and abandon the ordering assumption.  We find that
$\PE$ depends on the network and $Q_{out}(T_{out})$ only, while $\A$
depends on the network and $Q_{in}(T_{in})$ only.  We can prove global
upper and (unlike in the general case) lower bounds on $\PE$ and $\A$.

\begin{assumption}[Unclustered Assumption]
  Given a sequence of networks $\{G_n\}$, we assume that $G_n$ has
  girth greater than $2n$.
\end{assumption}

This assumption means that $B_d(u)$ chosen from any $G_n$ with $n \geq
d$ must be cycle free.  In particular, there is no alternate path
between a node and a neighbor.  It was this complication that forced
the use of the ordering assumption earlier, and since the complication
no longer exists, we drop the ordering assumption.  The unclustered
assumption will also allow us to use $\psi_{in}(V)$ rather than
$\phi_{in}(V,\vec{S})$.  Thus we only require the marginal probability
of the set of nodes $V$ not to be infected to satisfy an inequality,
rather than the inequality be satisfied for every possible
set of susceptibilities.  We must bear in mind that knowing $T_{in}$ or
$T_{out}$ no longer uniquely determines $\I$ or $\Sus$.

\begin{lemma}
  Let the sequence $\{G_n\}$ satisfy the unclustered assumption with
  sequential convergence of statistics.  Take $P_1(\I)$, $P_1(\Sus)$
  and $T_1(\I,\Sus)$.  Let $\psi_{in,1}(V)$ be as in
  equation~\eqref{eqn:psiV}.  Similarly take $P_2(\I)$, $P_2(\Sus)$,
  and $T_2(\I,\Sus)$ with corresponding $\psi_{in,2}(V)$.  If
  $\psi_{in,1}(V) \leq \psi_{in,2}(V)$ then $\PE_1 \geq \PE_2$.
  \label{lem:general_unclustered}
\end{lemma}

\begin{proof}
This proof is similar to that of Lemma~\ref{lem:general}.

Let $d \geq 0$ be given.  Take $G_n$, $n \geq d$.
Choose a node $u$ from $G_n$ and partition the nodes of $G_n$ into
$\{u\}$, $U_1$, and $U_2$.  To the nodes in $U_1$ we assign $\I$ from
$P_1(\I)$ and to the nodes of $U_2$ we assign $\I$ from $P_2(\I)$.  To
each node $w$ (including $u$), we assign two susceptibilities,
$\Sus_{w,1}$ and $\Sus_{w,2}$ such that $\Sus_{w,1}$ comes from
$P_1(\Sus)$ and $\Sus_{w,2}$ comes from $P_2(\Sus)$.

We create a partial EPN $\mathcal{E}$ as follows.  For each $v \in
U_1$, we assign edges $(v,w)$ using $T_1(\I_v, \Sus_{w,1})$, and for
$v \in U_2$ we assign them using $T_2(\I_v,\Sus_{w,2})$.  We do not
yet assign edges from $u$ (but edges may point to $u$).  Consider any
$u'$ not in $H_{in}(d)$ which would join $H_{in}(d)$ if an edge was added from $u$ to any $v \in V$.  By assumption, $\psi_{in,1}(V) \leq \psi_{in,2}(V)$ and
so the probability is greatest if $\I_u$ is chosen from $P_1(\I)$.  It
follows that $\PE_1 \geq \PE_2$.  This completes the proof.
\qed
\end{proof}

This proof can be modified to work on clustered networks without the
ordering assumption, so Lemma~\ref{lem:general} does not require the
ordering assumption.  However, the proof is more technical and
provides little additional insight, particularly because the main
results following from Lemma~\ref{lem:general} do require the
ordering assumption.


\begin{theorem}
  Let the sequence $\{G_n\}$ satisfy the unclustered
  assumption with sequential convergence of local statistics.  Let
  $Q_{in}(T_{in})$ be fixed.  Then $\A$ is fixed.
  \label{thm:attack_unclustered}
\end{theorem}

\begin{proof}
We follow the technique used to prove $\A$ is larger for one distribution than the other in Lemma~\ref{lem:general}.  However, in following that proof, the lack of clustering means $|V|=1$.  Since for any distribution $\psi(V)=1-\ave{T}$ when $|V|=1$, all distributions must give the same $\A$, and the proof is finished.
\qed
\end{proof}

\begin{theorem}
  If the assumptions of Theorem~\ref{thm:attack_unclustered} hold
  except that $Q_{out}$ is fixed rather than $Q_{in}$, then $\PE$ is
  fixed.
  \label{thm:prob_unclustered}
\end{theorem}
\begin{proof}
This follows immediately from Lemma~\ref{lem:interchange} and Theorem~\ref{thm:attack_unclustered}.  
\qed
\end{proof}

\begin{theorem}
  \label{thm:upper_prob_unclustered}
  Let $Q_{in}$ be given.  Assume that $\{G_n\}$ satisfies the
  unclustered assumption with sequential convergence of statistics.
  $\PE$ is maximized when $T(\I,\Sus) = T_{in}(\Sus)$.
\end{theorem}
Although this result is analogous to Theorem~\ref{thm:OAupper}, the proof is fundamentally altered because we no longer have the ordering assumption.

\begin{proof}
We first note that if $T(\I,\Sus) = T_{in}(\Sus)$, then $T_{out}=
\ave{T}$ for all nodes.

Now consider an arbitrary function $T(\I,\Sus)$ with $P(\I)$ and
$P(\Sus)$ to satisfy $Q_{in}(T_{in})$.  The function
$(1-T_{out})^{|V|}$ in equation~\eqref{eqn:psiV} is convex, so by
Jensen's inequality $\psi_{in}$ is minimized by $T_{out}= \ave{T}$.
Lemma~\ref{lem:general_unclustered} completes the proof.
\qed
\end{proof}

\begin{theorem}
\label{thm:lower_prob_unclustered}
  Let $Q_{in}$ be given.  Assume $\{G_n\}$ satisfies the unclustered assumption with sequential convergence of statistics.  $\PE$ is minimized when $\I$ is chosen
  uniformly from $[0,1]$ and
\[
T(\I,\Sus) = \begin{cases} 0 & \I>T_{in}(\Sus)\\ 1 & \I< T_{in}(\Sus)  \, .
\end{cases}
\]
\end{theorem}

\begin{proof}
  Following the proof of Theorem~\ref{thm:OAlower}, we may show that
  $\phi_{in}$ is maximized (subject to $Q_{in}$) exactly when these
  assumptions hold.  Thus from equation~\eqref{eqn:psiphi} $\psi_{in}$
  is also maximized when these assumptions hold.
  Lemma~\ref{lem:general_unclustered} completes the proof.  \qed
\end{proof}

\begin{theorem}
  Let $Q_{out}$ be given.  Assume $\{G_n\}$ satisfies the unclustered assumption with sequential convergence of statistics.
\begin{itemize}
\item $\A$ is maximized when $T(\I,\Sus) = T_{out}(\I)$.
\item $\A$ is minimized when $\Sus$ is chosen uniformly from $[0,1]$ and 
\[
T(\I,\Sus) = \begin{cases} 0 & \Sus>T_{out}(\I)\\ 1 & \Sus< T_{out}(\I) \, .\end{cases}
\]
\end{itemize}
\end{theorem}
\begin{proof}
  This follows from Theorems~\ref{thm:upper_prob_unclustered}
  and~\ref{thm:lower_prob_unclustered} with Lemma~\ref{lem:interchange}.
\qed
\end{proof}

Before proving our final result, we introduce a lemma.
\begin{lemma}
  Let $f(x)$ be a convex function on $[0,1]$ and $\rho(x)$ be a
  probability density function on $[0,1]$, with expected value
$\rho_0$.  Then
\[
\int f(x) \rho(x) dx \leq (1-\rho_0)f(0) + \rho_0 f(1) \, .
\]
\label{lem:inverseJensen}
\end{lemma}
\begin{proof}
The definition of convexity gives
\[
f(x) \leq (1-x) f(0) + x f(1) \, .
\]
Thus
\[
\int f(x) \rho(x) dx \leq \int [(1-x)f(0)\rho(x) + x f(1)\rho(x) ] dx \leq (1-\rho_0) f(0) + \rho_0 f(1) \, .
\]
\qed
\end{proof}

\begin{theorem}
\label{thm:unclustered_global}
Let $\{G_n\}$ be a sequence of networks satisfying the unclustered
assumption with sequential convergence of statistics.  Assume that
$\ave{T}$ is given:
\begin{itemize}
\item   The global upper bound for both $\PE$ and $\A$ occurs when $T_{uv} =
  \ave{T}$ for all pairs of neighbors.
\item The global lower bound for $\PE$ occurs when $Q_{out}(T_{out}) =
  \ave{T} \delta(T_{out}-1) + (1-\ave{T}) \delta(T_{out})$.
\item The global lower bound for $\A$ occurs when $Q_{in}(T_{in}) =
  \ave{T} \delta(T_{in}-1) + (1-\ave{T}) \delta(T_{in})$.
\end{itemize}
\end{theorem}
\begin{proof}
  The proof of the upper bound is identical to that of
  Theorem~\ref{thm:OAglobal_upper}.

  We prove the lower bound for $\PE$.  The lower bound for $\A$ 
  follows from Lemma~\ref{lem:interchange}.

  We have $\psi_{in}(V) = \int (1-T_{out})^{|V|} Q_{out}(T_{out})
  dT_{out}$.  We now seek to find $Q_{out}$ which maximizes $\psi_{in}$
  in order to apply Lemma~\ref{lem:general_unclustered}.

  Since $(1-T_{out})^{|V|}$ is a convex function, we may apply
  Lemma~\ref{lem:inverseJensen} with $Q_{out}$ playing the role of
  $\rho$.  The maximum occurs when $Q_{out}(T_{out}) = \ave{T}
  \delta(T_{out}-1) + (1-\ave{T}) \delta(T_{out})$ and so
  Lemma~\ref{lem:general_unclustered} finishes the proof.
\qed
\end{proof}

Although the upper bound for both $\PE$ and $\A$ occurs when
$T_{uv}=\ave{T}$ for all pairs, our earlier results show that for
unclustered networks $\PE$ depends only on $Q_{out}(T_{out})$ and the
network, and so as long as $T_{out}(u)=\ave{T}$ for all nodes $u$, we
achieve the upper bound on $\PE$ (but not on $\A$).  Symmetrically, if
$T_{in}(v)=\ave{T}$ for all nodes $v$, we achieve the upper bound on
$\A$.

Note that the lower bound for $\A$ requires that $T_{out}(u) =
\ave{T}$ for all $u$, and so the population is homogeneously
infectious.  It follows from Theorem~\ref{thm:lower_prob_unclustered}
that $\PE$ is then maximal.  Similarly the lower bound for $\PE$
requires that $\A$ be maximal.

\subsection{Discussion}

The results of this section generalize those
of~\cite{miller:heterogeneity} which considered the special case of
Molloy--Reed networks.  These results prove that the same scenarios
give upper and lower bounds in unclustered networks with a wide range
of correlations including assortative or disassortative mixing (high
degree nodes preferentially joining with high or low degree nodes
respectively), or longer range correlations.  Although we proved these
under the assumption that no short cycles exist, the results remain
useful in networks with either few short cycles, or for situations in
which the transmissibility is low enough that the short cycles are
only rarely followed.

Because of the lack of short cycles, the ordering assumption is not
needed.  This means that our results apply to a much wider class of
disease transmission mechanisms, but at the cost of restricting the
network.  Again we find that which conditions give the upper or lower
bound is network-independent.  The amount of variation there is between
these bounds, however, is network-dependent.

The main distinction from clustered networks is that $\PE$ depends
only on the network structure and $Q_{out}(T_{out})$.  That is, $\PE$
is independent of $Q_{in}(T_{in})$.  Similarly $\A$ depends only on
the structure and $Q_{in}(T_{in})$.  We note that unless the effect of
clustering is very large, the dependence of $\PE$ on
in-transmissibility and $\A$ on out-transmissibility will be weak.
Curiously the global lower bound for $\A$ found in
Theorem~\ref{thm:unclustered_global} requires that $Q_{out}(T_{out}) =
\delta(T_{out}-\ave{T})$, and so the population is homogeneously
susceptible.  It follows from Theorem~\ref{thm:lower_prob_unclustered}
that $\PE$ is then maximal.  Similarly the lower bound for $\PE$
requires that $\A$ be maximal.  This has important implications for
policy design because strategies to reduce $T$ tend to have a
heterogeneous impact on either $\Sus$ or $\I$.

\section{Conclusions}
\label{sec:conclusions}
We have extended earlier work on the effect of heterogeneity in
infectiousness on the spread of epidemics through networks.  Our
extensions allow for heterogeneity in susceptibility as well.  Many of
the results are similar.  In general we find that the size and
probability of epidemics are reduced if the population is more
heterogeneous.  Unfortunately, increasing heterogeneity in
susceptibility restricts the level of heterogeneity possible in
infectiousness.  In the extreme case where susceptibility is maximally
heterogeneous, infectiousness must be homogeneous.  Perhaps
surprisingly, we have found that the distributions leading to upper
and lower bounds on $\PE$ and $\A$ are network-independent.

Early in an outbreak, it is likely that we may gain some information
about $Q_{out}(T_{out})$.  For example, in the early stages of the
SARS epidemic, it was known that a number of people were highly
infectious, while the rest were only mildly infectious and so
$Q_{out}(T_{out})$ was highly heterogeneous.  However, there was
little information on $Q_{in}$.  Once given the distribution of
$Q_{out}$, the results here show which distributions of $Q_{in}$ give
the largest or smallest $\PE$ and $\A$.  Our results further suggest
that the distribution of infectiousness found for SARS is consistent
with a low epidemic probability.  It is difficult to extrapolate from
observations what the sizes would have been without the interventions
put into place, but the fact that a number of isolated cases occurred
throughout the world without sparking local epidemics suggest that the
probability of an epidemic from each introduction was low, consistent
with our predictions.

Our results further suggest that in order to prevent an epidemic, it
is best to take measures that will have a heterogeneous impact on
infectiousness, but in order to affect the size of an epidemic, it is
best to take measures that will have a heterogeneous impact on
susceptibility.  In terms of actual interventions, we compare two
strategies aimed at controlling a disease which is initially spreading
with homogeneous $T$: in the first we devote resources to vaccinating
half of the population, while in the second we devote them to
identifying and removing half of the infected population.  Both
strategies reduce $\ave{T}$ by a factor of $2$.  In the first, the
susceptibility is highly heterogeneous, but the probability an
infected node infects a randomly chosen neighbor has simply gone down
by a factor of $2$, and so it remains homogeneous. Assuming that the
unclustered approximation is valid, this maximizes the impact on size,
but the impact on probability is minimized.  In contrast, the second
strategy maximizes the impact on probability, but minimizes the impact
on size.

\section*{Acknowledgments}
This work was supported in part by the Division of Mathematical
Modeling at the UBC CDC and by DOE at LANL under Contract
DE-AC52-06NA25396 and the DOE Office of ASCR program in Applied
Mathematical Sciences.

\bibliographystyle{apt} 
\bibliography{Probsizebounds}

\end{document}